# A Review for Japanese auroral records on the three extreme space weather events around the International Geophysical Year (1957 – 1958)

Hisashi Hayakawa (1 – 4)*, Yusuke Ebihara (5 – 6), Hidetoshi Hata (7 – 8)

(1) Institute for Space-Earth Environmental Research, Nagoya University, Nagoya, 4648601, Japan

(2) Institute for Advanced Researches, Nagoya University, Nagoya, 4648601, Japan

(3) RAL Space, Rutherford Appleton Laboratory, Science and Technology Facilities Council, Harwell Campus, Didcot, OX11 0QX, UK

(4) Nishina Centre, Riken, Wako, 3510198, Japan

(5) Research Institute for Sustainable Humanosphere, Kyoto University, Uji, 6110011, Japan

(6) Unit of Synergetic Studies for Space, Kyoto University, Kyoto, 6068306, Japan

(7) Kisomachi Junior High School, 4110 Shinkai, Kisomachi, Kiso-gun, Nagano, 3970002, Japan

(8) Kiso Observatory, Institute of Astronomy, School of Science, the University of Tokyo, 10762-30, Mitake, Kiso-machi, Kiso-gun, Nagano 3970101, Japan

* hisashi@nagoya-u.jp

**Abstract**

Solar Cycle 19 was probably the greatest solar cycle over the last four centuries and significantly disrupted the solar-terrestrial environments with a number of solar eruptions and resultant geomagnetic storms. At its peak, the International Geophysical Year (IGY: 1957 – 1958) was organised by international collaborations and benefitted scientific developments, capturing multiple unique extreme space weather events including the third and fourth greatest geomagnetic storms in the space age. In this article, we review and analyse original records of Japanese auroral observations around the IGY. These observations were organised by Masaaki Huruhata in collaboration with professional observatories and citizen contributors. We have digitised and documented these source documents, which comprise significant auroral displays in March 1957 (minimum Dst = −255 nT), September 1957 (minimum Dst = −427 nT), and February 1958 (minimum Dst = −426 nT). These records allow us to visualise temporal and spatial evolutions of these auroral displays, reconstruct their equatorward auroral boundaries down to 41.4°, 38.3°, and 33.3° in invariant latitudes, and contextualise their occurrences following contemporary geomagnetic disturbances. Our results have been compared with significant auroral displays during other extreme space weather events. These aurorae generally showed reddish colourations occasionally with yellowish rays. Their colourations are attributed to reddish oxygen emission and its mixture with greenish oxygen emission. Overall, these archival records provide the references for future discussions on the auroral activities during the uniquely intense and extreme space weather events.

**Keywords**

Space weather, aurorae, geomagnetic storms, Solar Cycle 19, International Geophysical Year

**1. Introduction**

Auroral visibility in mid- to low-latitude regions shows the evolution of geomagnetic storms, which originate from intense solar eruptions and the resulting interplanetary coronal mass ejections (ICMEs) (Gonzalez *et al*., 1994; Daglis, 2006). This was particularly the case in the space age, as exemplified with the extreme geomagnetic storms in March 1989 and February 1958, which recorded the greatest and fourth greatest geomagnetic disturbances in the disturbance storm-time (Dst) index since the International Geophysical Year (IGY), extended auroral visibility down to Mexico, and caused severe space weather effects such as blackouts and power system effects (Allen *et al*., 1989; Rich and Denig, 1992; Boteler *et al*., 1998; Silverman, 2006; Lanzerotti, 2017; Boteler, 2019; Knipp *et al*., 2021). Analyses of such great auroral displays are more than just a scientific concern, as historical evidence shows geomagnetic superstorms and significant auroral displays in the long term (Tsurutani *et al*., 2003; Cliver and Dietrich, 2013; Hayakawa *et al*., 2019b; Knipp *et al*., 2021) and their potential impacts have been considered even catastrophic to the modern technological infrastructure (Lanzerotti, 2017; Baker *et al*., 2018; Riley *et al*., 2018).

Such solar eruptions frequently occurred in maxima to declining phases in enhanced solar cycles (Lefèvre *et al*., 2016; Owens *et al*., 2021). In this context, Solar Cycle 19 is considered as the greatest solar cycle since 1610 (Clette *et al*., 2014; Hathaway, 2015; Clette and Lefèvre, 2016; Muñoz-Jaramillo and Vaquero, 2019). In the International Sunspot Number, this solar cycle spanned from April 1954 to October 1964 and peaked in October 1957 (359.4) with respect to the monthly mean and in March 1958 (285.0) with respect to the smoothed monthly mean (Clette *et al*., 2014; Clette and Lefèvre, 2016). The sun was notably eruptive in this cycle, launched numerous interplanetary coronal mass ejections (ICMEs) and solar energetic particles, and triggered a number of extreme space weather events (Cliver and Crooker, 1993; Rishbeth *et al.*, 2009; Lefèvre *et al.*, 2016; Cliver *et al.*, 2020; Usoskin *et al.*, 2020a, 2020b).

Solar-terrestrial environments were significantly disturbed during this solar cycle. In principle, extreme geomagnetic storms are rare despite their significant impacts on the technological infrastructure of human civilisation (Lanzerotti, 2017; Baker *et al.*, 2018; Riley *et al.*, 2018). Only 5 and 39 geomagnetic storms exceeded the thresholds of the minimum Dst ≤ −400 nT and ≤ −250 nT, respectively, within the standard Dst index since 1957 (WDC for Geomagnetism at Kyoto *et al.*, 2015; Riley *et al.*, 2018; Stanislawska *et al.*, 2018; Meng *et al.*, 2019). Solar Cycle 19 accommodates 3 of the 5 aforementioned geomagnetic storms (minimum Dst ≤ −400 nT) and 14 of the 39 aforementioned geomagnetic storms (minimum Dst ≤ −250 nT), even though its ascending phase was overlooked in the standard Dst index (WDC for Geomagnetism at Kyoto *et al.*, 2015). Such concentrations significantly distinguish Solar Cycle 19 from the other solar cycles from 1957 onward (Riley *et al.*, 2018). In Solar Cycle 22, only 1 geomagnetic storm exceeded the threshold of minimum Dst ≤ −400 nT and 9 storms exceeded the threshold of minimum Dst ≤ −250 nT, whereas it also hosted the greatest geomagnetic storm (the Hydroquebec superstorm on 13/14 March 1989) since the IGY (Allen *et al.*, 1989; Boteler, 2019). In Solar Cycle 23, only 1 geomagnetic storm exceeded the threshold of minimum Dst ≤ −400 nT and 10 storms exceeded the threshold of minimum Dst ≤ −250 nT.

During these extreme geomagnetic storms, the equatorward boundaries of the auroral *oval* and the auroral *visibility* extended toward the mid to low magnetic latitudes (Vallance Jones, 1992; Silverman, 2006), implying their empirical correlation with intensities of the associated geomagnetic storms (Akasofu and Chapman, 1963; Akasofu, 1964), which was later established by additional satellite data (Yokoyama *et al.*, 1998; Blake *et al.*, 2021). The IGY was organised around this maximum (1957 – 1958), formed a benchmark international scientific collaboration within the Cold War, and allowed for the elucidation of geoscience and creation of the system of World Data Centres (WDCs) (Odishaw, 1958, 1959; Sullivan, 1961). Japanese scientists took part in these international collaborations contributions in several fields including the auroral observations (Huruhata, 1958, 1960; Hirosaka, 1958).

Its legacies, including the WDC system, have benefitted modern science for more than six decades (Baker *et al.*, 2011; Lanzerotti and Baker, 2018). These geomagnetic storms significantly impacted the contemporary technological infrastructure and triggered scientific discussions on space weather hazards (Boteler *et al*., 1998; Lanzerotti, 2017). Since then, analyses of such extreme geomagnetic storms have increased in significance, as human civilisation has accelerated the dependency on the technological infrastructure and has in turn become significantly more sensitive to extreme geomagnetic storms (Lanzerotti, 2017; Baker and Lanzerotti, 2017; Riley *et al*., 2018; Balan *et al*., 2019). However, the scarcity of such extreme geomagnetic storms has made these individual cases rather unique. The problem has been further compounded owing to the limited number of observations that were available during these storms time. As such, it is important to analyse the contemporary observational records in the modern viewpoints. However, the IGY storms were insufficiently documented except for the February 1958 storm (Huruhata, 1960; Stanislawska *et al*., 2018), which has been retrospectively analysed and highlighted with spectroscopic observations (Saito *et al*., 1994; Kataoka *et al*., 2019a) and visual observations (Vallance Jones, 1992; Nakazawa, 1999; Silverman, 2006; Ninomiya, 2013; Lanzerotti and Baker, 2018; Kataoka *et al*., 2019b), as well as comparison with other extreme space weather events (Cliver and Svalgaard, 2004; Knipp *et al*., 2021). In this context, we review and analyse original Japanese records of visual auroral observations for three extreme geomagnetic storms around the IGY (March 1957, September 1957, and February 1958), whose intensities ranked 36th, 3rd, and 4th in the Dst index (Huruhata, 1960; WDC for Geomagnetism at Kyoto *et al*., 2015; Meng *et al*., 2019). We clarify the source documentations for these auroral records and reconstruct the spatial and temporal evolution of the auroral oval in the Japanese sector.

## 2. Data and Methods

Masaaki Huruhata (1912 – 1988) oversaw the Japanese contributions on aurorae and airglows in Antarctica and Japanese Islands during the IGY (Huruhata, 1956, 1957, 1960). Huruhata (1957) organised and called for systematic auroral observations in Japan, following discussions in the CASGI (*Comité spécial de l'année géophysique internationale*) working group and expecting potential auroral visibility in the low- to mid-latitude area (Huruhata, 1957; Chapman, 1957; Nicolet, 1959, p. 517). He requested auroral observations among meteorological observatories as well as citizen contributors to improve the geographical coverage of the planned auroral observations. Huruhata (1957) requested details on their morphology, brightness in the International Brightness Coefficient, temporal and spatial evolutions in a radar chart, and images captured using cameras.

As the first director of WDC for Airglow at Tokyo Astronomical Observatory, he gathered

contributions from meteorological stations across Japan including those of the Japan Meteorological Agency (JMA), eight groups of citizen astronomers, and eight airglow stations (Huruhata, 1960), including the earliest auroral images in Japan (JMA, 1958a, p. 29). These contributions have been collected in at least two institutes: the JMA and Tokyo Astronomical Observatory. The JMA collected individual reports from their local meteorological offices and published them in *Geophysical Review* with selected radar charts (JMA, 1957a, 1957b, 1957c, 1958a, 1958b). The JMA local meteorological offices recorded these auroral observations in their original daily ledgers as well. These ledgers are located in the archives of each local meteorological office and occasionally provide unique details that are not documented in the publications in *Geophysical Review*.

Huruhata's own collections were located in Tokyo Astronomical Observatory, which is currently known as the National Astronomical Observatory of Japan (NAOJ). These visual records remained mostly unpublished, except for the photometric observations during the February 1958 storm (Huruhata, 1958, 1960; Hirosaka, 1960; Kakioka Observatory, 1969). While most of their original records were abandoned for the room moving of WDC for Airglow at the NAOJ, certain graphical records have been preserved in the NAOJ Mitaka Library, as an archival collection entitled 'Sketches for aurorae that occurred during the International Geophysical Year', attributed to Kyoko Tanaka of the WDC for Airglow. Fortunately, before the record abandonments upon the WDC room-moving, Hidetoshi Hata (one of the authors of this article) communicated with Kyoko Tanaka and managed to salvage their digital images and store them into the CD ROM, as explained in Hata (2000). Their digital images are currently preserved in a CD ROM "Kiso Schmidt Astronomical Image Collection", which was compiled by Hidetoshi Hata, Yoshikazu Nakada, and Tsutomu Aoki and preserved at Kiso Observatory, Institute of Astronomy, School of Science of the University of Tokyo.

In this study, we reviewed the JMA publications in *Geophysical Review* (JMA, 1957a, 1957b, 1958a, 1958b) and daily lodgers in the individual local meteorological offices, extracted the visual records in the CD ROM at Kiso Observatory, and took pictures of the archival collection in the NAOJ Mitaka Library. We reviewed and compiled these auroral reports to extract their observational sites, visibility durations, time offset with the Universal Time (UT), reported directions, colourations, maximal altitudes, and source documents. We also computed the magnetic latitude (MLAT) for each observational site, using the IGRF-12 model (Thébault *et al.*, 2015). These records provide source data for four of the five notable auroral displays in Huruhata's summary (Huruhata, 1960), except for the one on 13 December 1958. We have analysed their descriptions, visualised their spatial and temporal evolutions, and contextualised these records using contemporary geomagnetic measurements.

## 3. Auroral Reports in March 1957

This observational network captured the first great auroral display upon the occurrence of an extreme geomagnetic storm (minimum Dst = − 255 nT) on 2/3 March 1957 (WDC for Geomagnetism at Kyoto *et al.*, 2015; Meng *et al.*, 2019). At the time, a great auroral display was extensively observed in Hokkaido and Tohoku Regions in Japan during 20 – 23 LT. Figure 1 shows their examples: a radar chart from Kuji and a drawing from Wassamu. These records, including the above two records (Figure 1), show reddish glows with occasional yellowish to whitish ray structure, and confirm auroral visibility at least from Rebun (35.3° MLAT) to Kuji (30.1° MLAT), as summarised in Figure 2. The weather was generally favourable for these observations, as there was little cloud cover in northern Japan, in contrast with western Japan, according to contemporary weather charts[1].

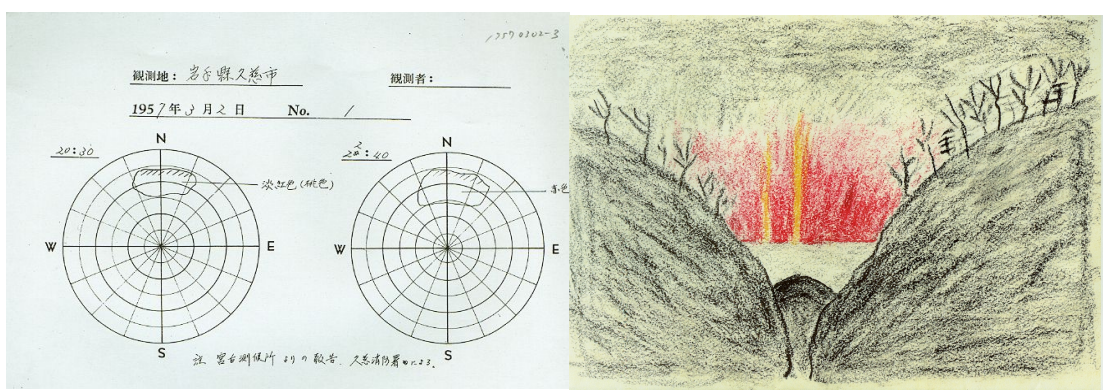

Figure 1: Japanese auroral records on 2 March 1957, including an auroral radar chart from Kuji (left) and an auroral drawing from Wassamu (right), reproduced courtesy of © Miyako Observatory, Kiichiro Sase, and © Kiso Observatory of the University of Tokyo.

---

[1] http://agora.ex.nii.ac.jp/digital-typhoon/weather-chart/index.html

Hayakawa et al. (2021) Japanese auroral records in the IGY (1957 – 1958)
*Geoscience Data Journal*, DOI: 10.1002/GDJ3.140

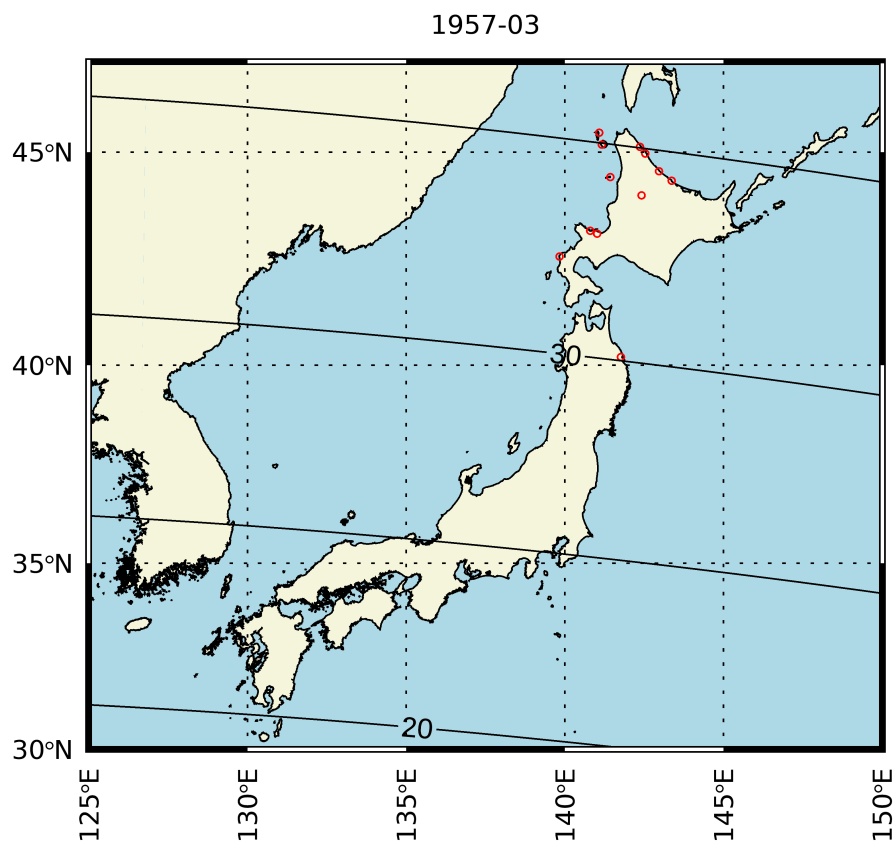

Figure 2: Geographical distributions of the auroral visibility during the geomagnetic storm on 2 March 1957. The contour lines indicate the magnetic latitudes (MLAT) at 20°, 25°, 30° and 35°.

We confirmed auroral displays with distinct ray structure down to Mochita (32.3° MLAT) and Wassamu (34.0° MLAT). As the auroral display extended up to 30° in elevation at Wassamu, we conservatively computed the equatorward boundary of the auroral oval as 40.6° ILAT (invariant latitude in the dipole magnetic field), in terms of the footprint of the magnetic field line along which the auroral electrons precipitated, assuming an auroral elevation of 400 km (Roach *et al.*, 1960; Ebihara *et al.*, 2017). At Kuji, the reddish glows extended to 45° at maximum elevation (Figure 1 (left)). The reddish glows observed at Kuji are rather monochromatic in colouration and hence a part of the reddish glows may be interpreted as stable auroral red (SAR) arcs (Kozyra *et al.*, 1997). Assuming that reddish glows observed at Kuji were fully auroral origin and have elevation of 400 km, we compute its equatorward boundary as 35.9° ILAT.

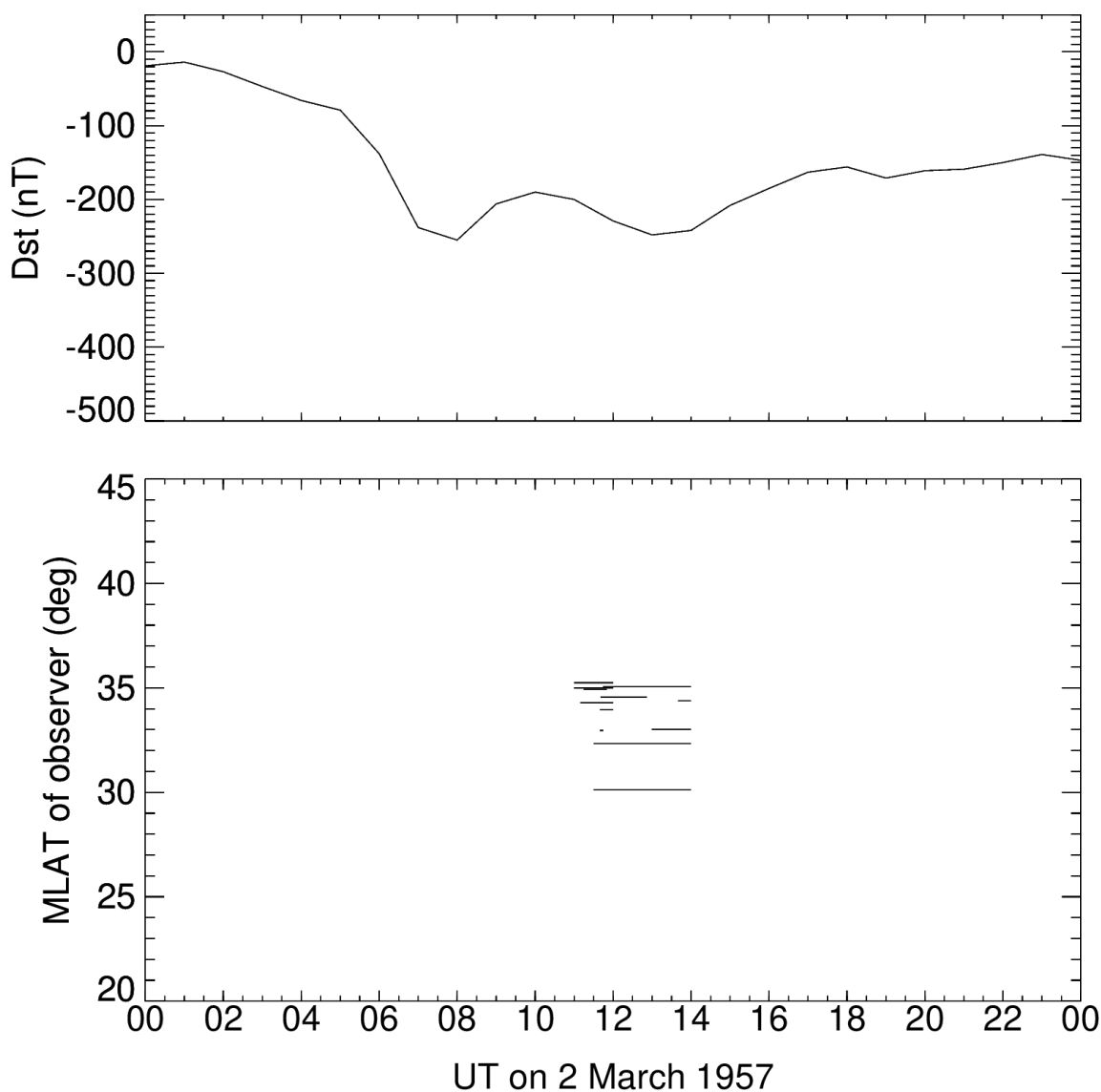

Figure 3: Temporal and spatial evolution of the auroral visibility on 2 March 1957 (lower panel), contextualised in temporal variation of the Dst index. In the lower panel, the Japanese local time (UT + 9 h) has been corrected to the UT.

Figure 3 contextualises temporal evolutions of these auroral visibilities upon the contemporary geomagnetic disturbance represented in the Dst index (WDC for Geomagnetism at Kyoto *et al.*, 2015). These auroral displays chronologically coincide with an extreme geomagnetic storm (minimum Dst = −255 nT), which peaked at 8 UT on 2 March. The auroral displays were reported during 20 – 23 LT in Japan (11 – 14 UT). They were located around the second peak in the recovery phase of this geomagnetic storm, especially when the Dst index dropped below < −200 nT after the initial short recovery.

### 4. Auroral Reports in September 1957

Another great auroral display was captured in the observational network on 13/14 September 1957, upon occurrence of the third largest geomagnetic storm (minimum Dst = −427 nT) in the Dst index (WDC for Geomagnetism at Kyoto *et al.*, 2015; Meng *et al.*, 2019). We identified four coloured auroral drawings in the NAOJ Mitaka Library, as exemplified in Figure 4. In addition, we identified visual auroral reports including radar charts in JMA (1957c) and Kiso Observatory, as exemplified in Figure 5. Figure 6 shows the geographical extent of the reported auroral visibility on 13/14 September, spanning from Wakkanai (35.3° MLAT) to Mori (31.9° MLAT). These observations are geographically confined in the western part of Hokkaido Island, mainly because the northern to eastern parts of Japan were mostly under cloud cover except for the western part of Hokkaido, according to contemporary weather charts.

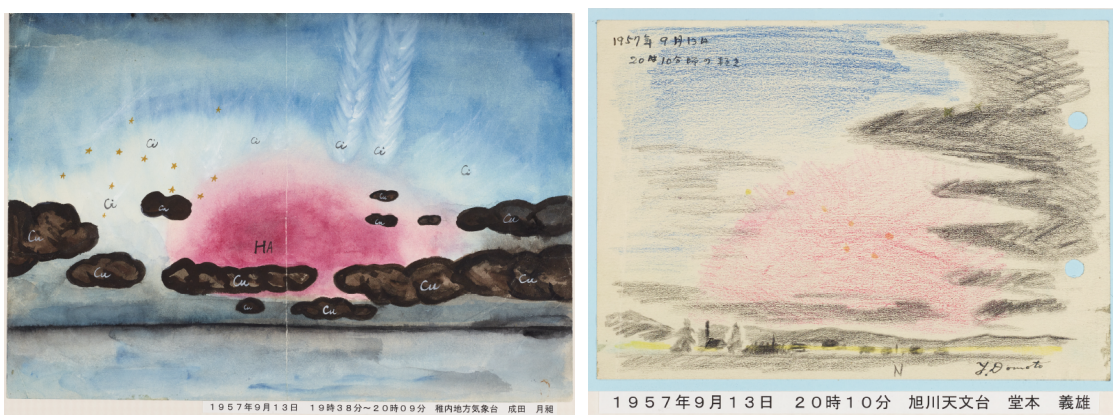

Figure 4: Japanese auroral drawings on 13 September 1957, including Narita's auroral drawing at Wakkanai and Domoto's auroral drawing at Asahikawa, reproduced courtesy of © Yoshio Domoto, © Tsukihisa Narita, and © the National Astronomical Observatory of Japan.

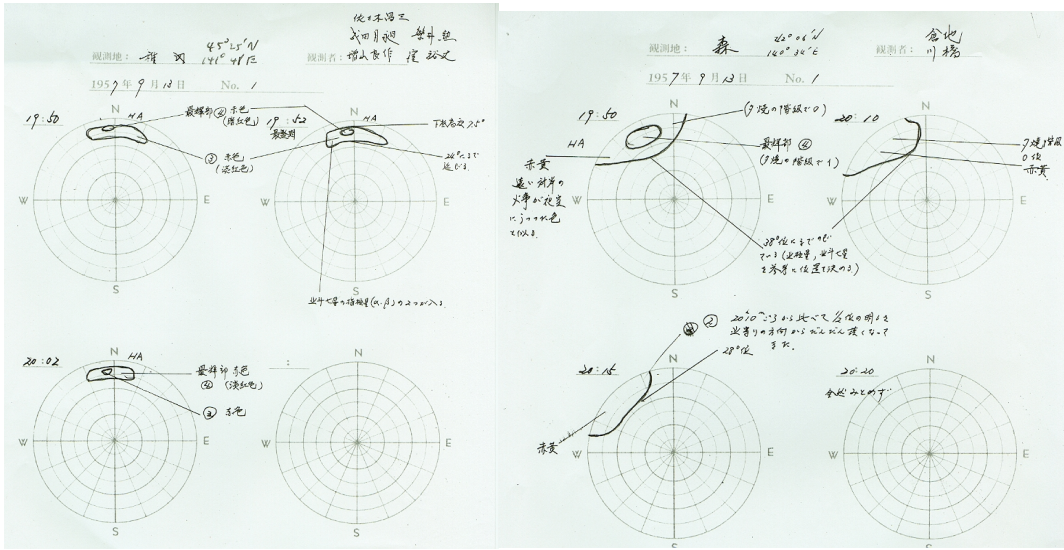

Figure 5: Auroral radar charts on 13 September 1957, reported from Wakkanai (left) and Mori (right), reproduced courtesy of © Narita Tsukihisa, © Kurachi and Kawahashi, and © Kiso Observatory of the University of Tokyo.

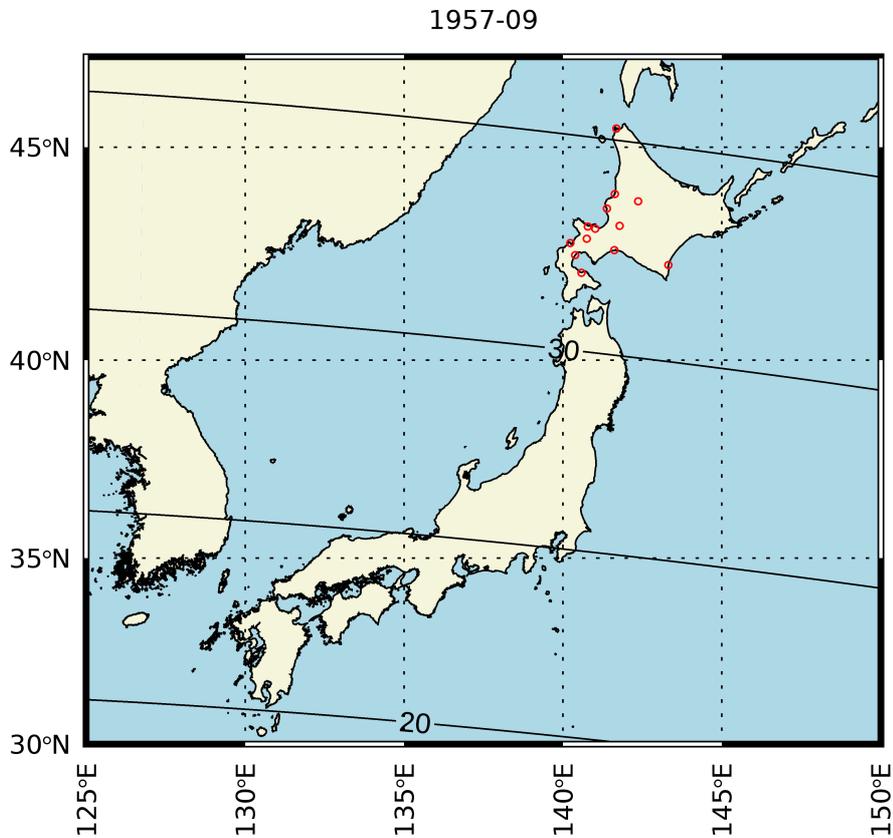

Figure 6: Geographical distributions of the auroral visibility during the geomagnetic storm on 13

September 1957.

These auroral displays were mostly reddish, while orange components were reported at Mori and Rumoi. Among these records, Mori was situated in the lowest MLAT (31.9° MLAT). Here, the auroral display was visible up to 38° in elevation (JMA, 1957c, p. 33). On this basis, we compute its equatorward auroral boundary as 38.3° ILAT, assuming the auroral elevation to be 400 km (Roach *et al.*, 1960; Ebihara *et al.*, 2017). Their colourations indicate that the reported aurorae were probably not SAR arc (Kozyra *et al.*, 1997).

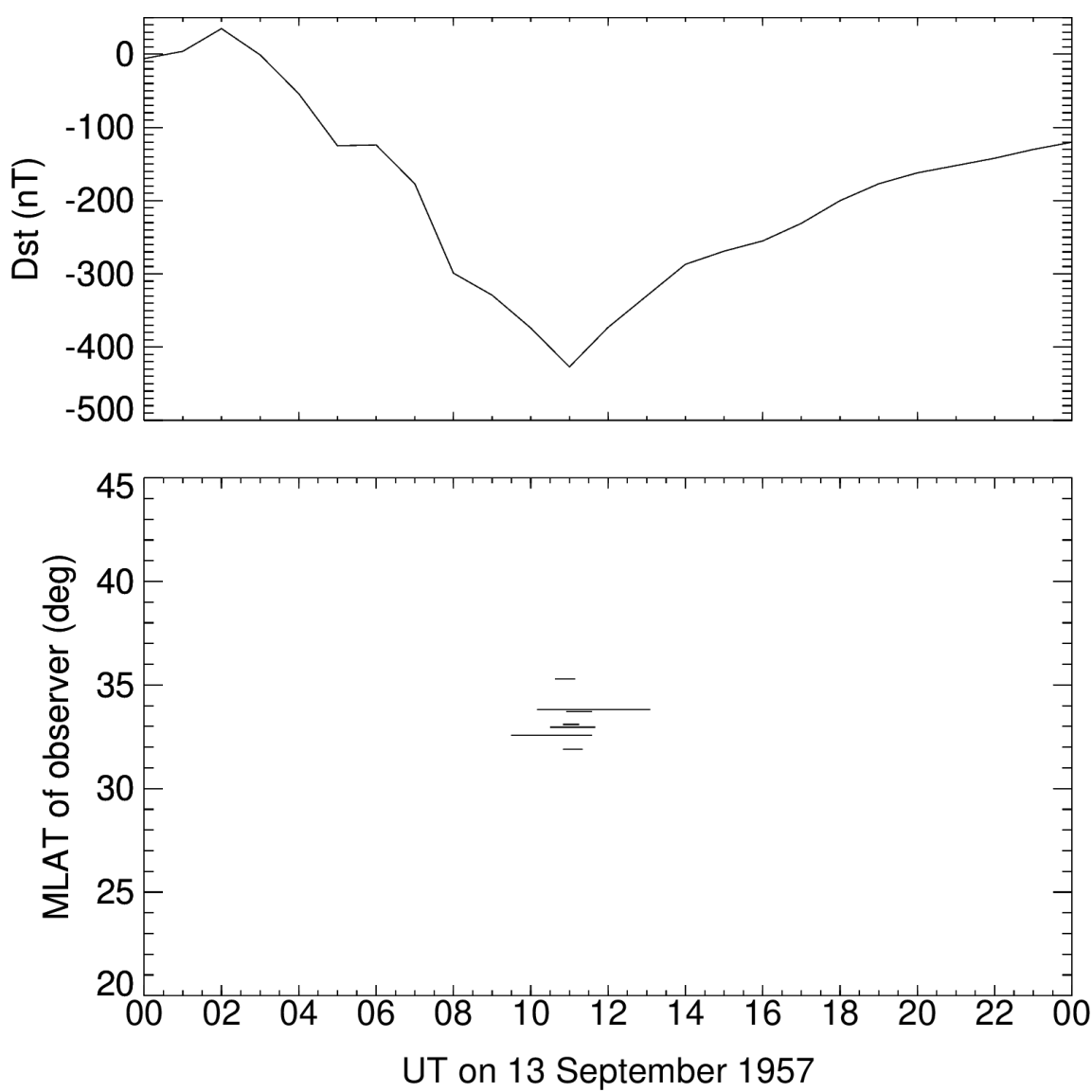

Figure 7: Temporal and spatial evolution of the auroral visibility on 2 March 1957 (lower panel), contextualised in temporal variation of the Dst index. In the lower panel, the Japanese local time

(UT + 9 h) has been corrected to the UT.

Figure 7 contextualises temporal evolutions of these auroral visibilities upon the contemporary geomagnetic disturbance represented in the Dst index (WDC for Geomagnetism at Kyoto *et al.*, 2015). These auroral reports chronologically coincided with the third greatest geomagnetic storm in the Dst index (minimum Dst = −427 nT), which peaked at 11 UT on 13 September. Their visibilities were reported during 18:30 – 22:05 LT in Japan (09:30 – 13:05 UT). Hence, they are located in the main phase to the early recovery phase of this storm, where the Dst index exceeded the threshold of Dst ≤ −300 nT.

## 5. Auroral Reports in February 1958

The great auroral display on 11/12 February 1958 is probably what has been most documented, discussed, and analysed among the great auroral displays and the extreme geomagnetic storms during/around the IGY in modern scientific literature (Vallance Jones, 1992; Cliver and Svalgaard, 2004; Lanzerotti and Baker, 2018; Kataoka *et al.*, 2019a, 2019b; Knipp *et al.*, 2021). At the time, the auroral display was reported down to Mexico City (29.3° MLAT) in the North American sector (Rivera-Terrezas and Gonzalez, 1964; Cliver and Svalgaard, 2004; Knipp *et al.*, 2021) and down to Aikawa and Niigata in the East Asian sector (Hikosaka, 1958; Huruhata, 1958; Saito *et al.*, 1994; Kataoka *et al.*, 2019a, 2019b), as documented in recent publications. Japanese articles have indicated the availability of additional auroral records even in lower MLATs such as Nagano, Kanto, and western Japan (Nakazawa, 1999; Ninomiya, 2013). These wide-range auroral visibilities benefitted from the Japanese weather condition at the time with limited cloud cover (Figure 2 of Ninomiya (2013)).

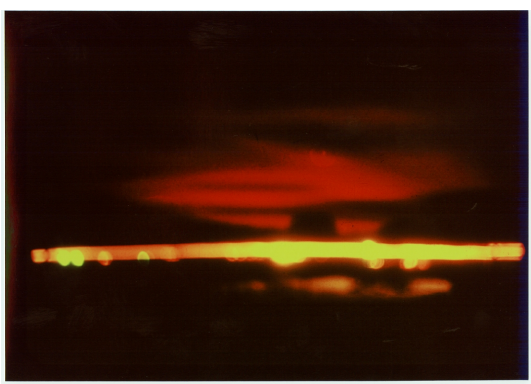

Figure 8: The possibly earliest known coloured auroral photograph in Japan captured at Shizunai on 11/12 February 1958, courtesy of © Setsuya Hasegawa and © Kiso Observatory of the University of Tokyo.

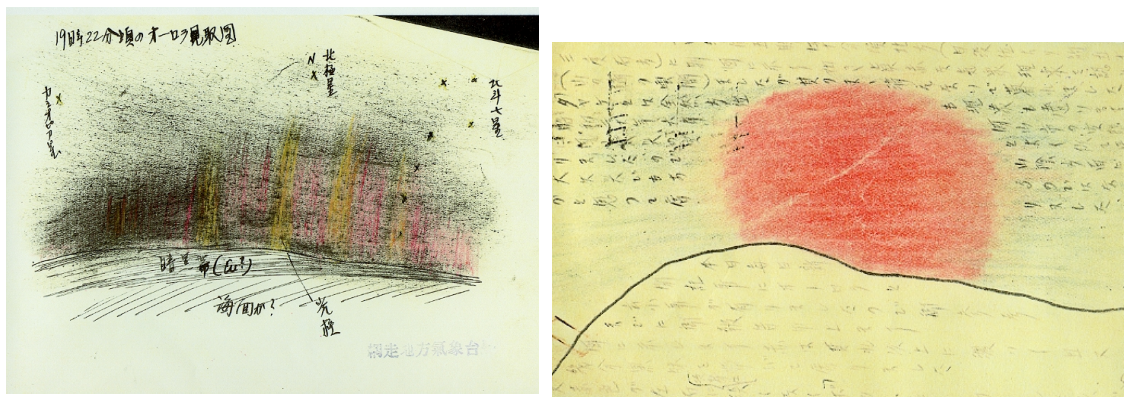

Figure 9: Auroral drawings reported from Abashiri Local Meteorological Office (left) and Fukuyama (right) on 11/12 February 1958, courtesy of © Abashiri Local Meteorological Office, © Yoshio Mimura, and © Kiso Observatory of the University of Tokyo.

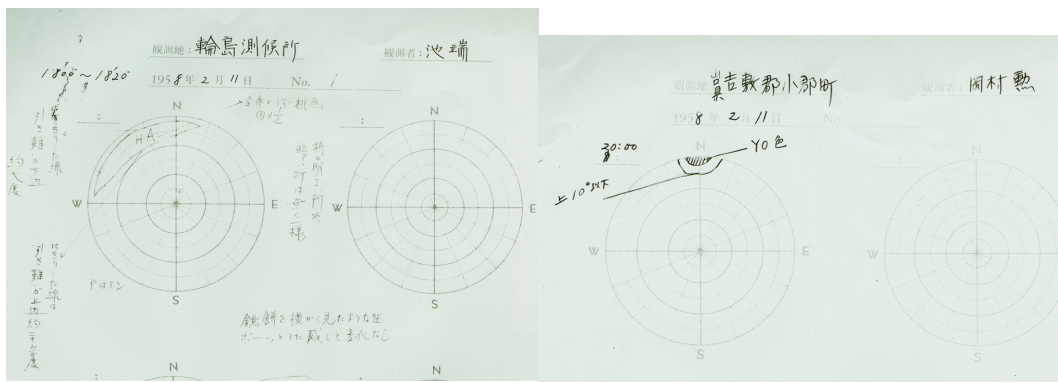

Figure 10: Auroral radar charts on 11/12 February 1958, reported from Wajima (left) and Ogori (right), courtesy of © Wajima Local Meteorological Office, © Isao Okamura, and © Kiso Observatory of the University of Tokyo.

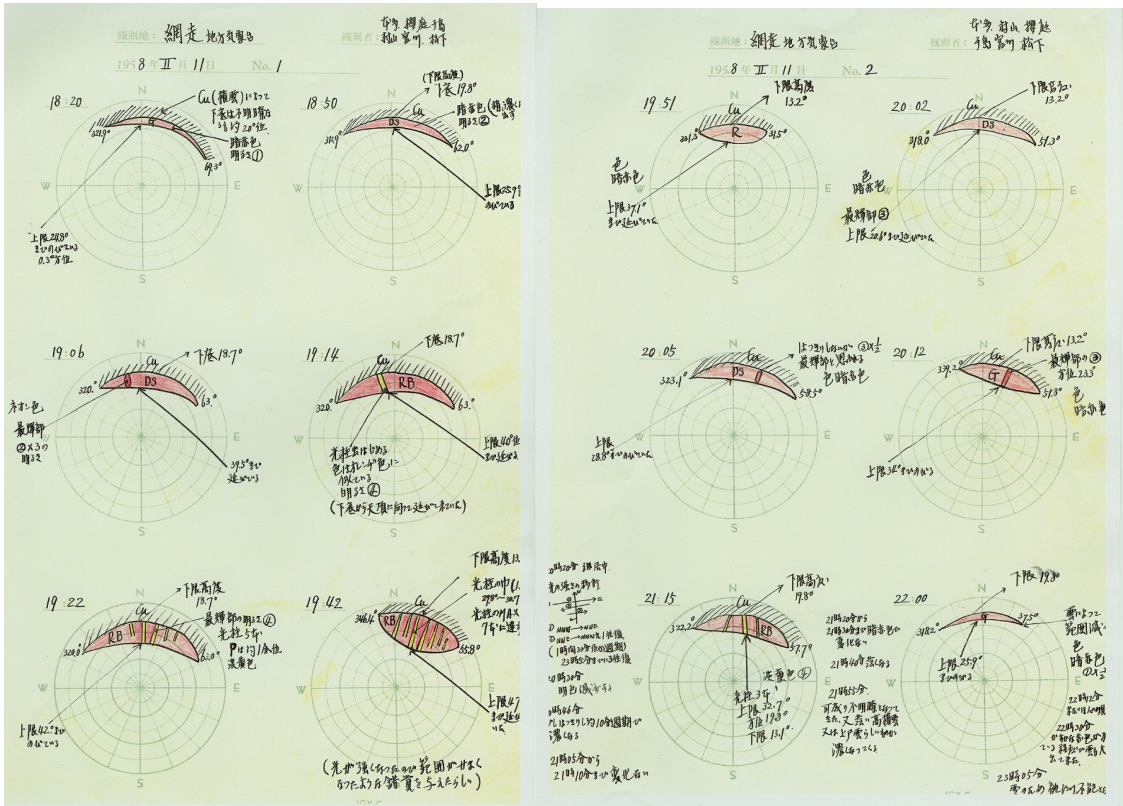

Figure 11: Colour auroral radar charts on 11/12 February 1958, reported from Abashiri Local Meteorological Office, showing the auroral temporal evolution from 18:20 to 22:00 LT, courtesy of © Abashiri Local Meteorological Office and © Kiso Observatory of the University of Tokyo.

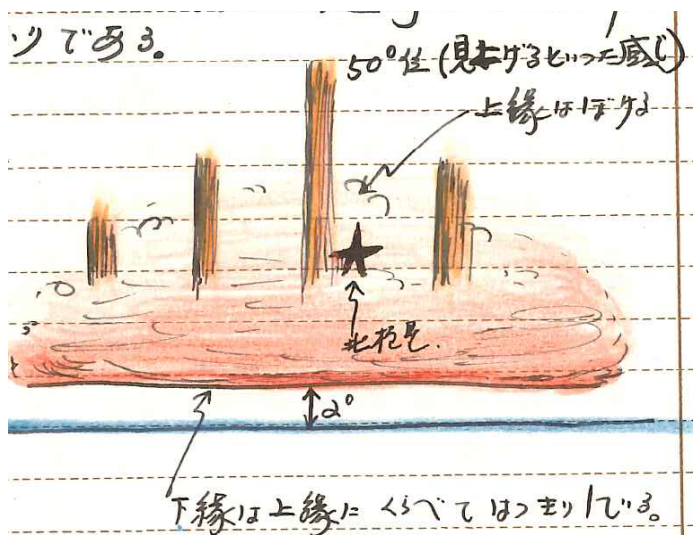

Figure 12: A sample auroral drawing in the daily ledger of Niigata Local Meteorological Office, courtesy of © Niigata Local Meteorological Office.

While the NAOJ Mitaka Library has only preserved Shigeru Kazama's auroral drawing (see Kataoka *et al.*, 2019b), Kiso Observatory has preserved copies of additional images, drawings, and radar charts, as exemplified in Figures 8 – 11. We located a coloured auroral photograph from Shizunai (32.3° MLAT; Figure 8), while the published auroral photographs at the time have all been illustrated without colouration (*e.g.*, JMA, 1958a; Kakioka Observatory, 1969). The image that we identified is probably the earliest coloured auroral photograph in Japan, as it features the first photographed aurora in Japanese history (JMA, 1958a, p. 29). Additionally, we located numerous auroral drawings (*e.g.*, Figure 9 and 12) and auroral radar charts (*e.g.*, Figures 10 – 11) in JMA (1958a, 1958b) and in the collections of Kiso Observatory and Niigata Local Meteorological Office.

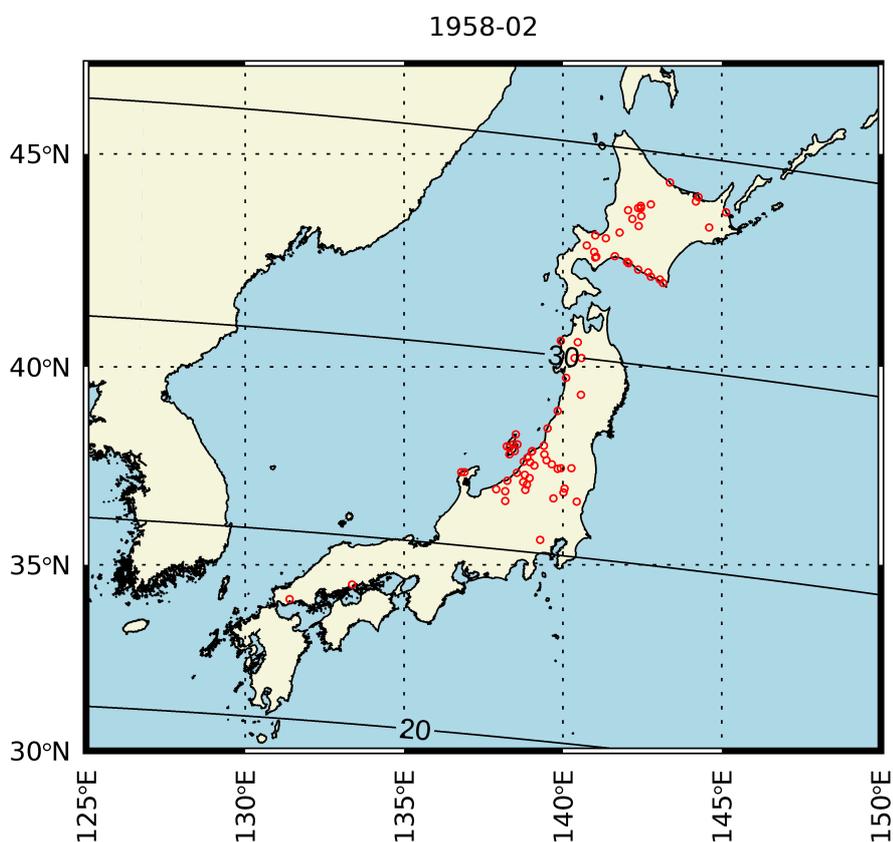

Figure 13: Geographical distributions of auroral visibility during the geomagnetic storm on 11 February 1958.

Figure 13 shows the geographical extent of the reported auroral visibility on 11/12 February, spanning down to Ogori (23.3° MLAT; Figure 10b) and Fukuyama (23.8° MLAT; Figure 9b). The reported auroral elevation of 10° at Ogori indicates the equatorward auroral boundary as 37.7° ILAT.

Furthermore, auroral records from Niigata (27.7° MLAT; Figure 12) and Wajima (26.9° MLAT; Figure 10a) reported spatial extents of up to 50° and 25° in their elevations (Figure 10a and 12). These records locate the equatorward auroral boundaries at 33.3° ILAT and 35.9° ILAT, respectively. Among these records, the auroral report from Niigata locates the equatorward auroral boundary at the lowest ILAT. As this record distinctly shows ray structure, it does not indicate a SAR arc but a regular auroral emission. Therefore, we located the equatorward boundary of the auroral oval during this storm at 33.3° ILAT, which is significantly more equatorward than the existing estimates of 38° – 40° based on the Aikawa report (Kataoka *et al.*, 2019a).

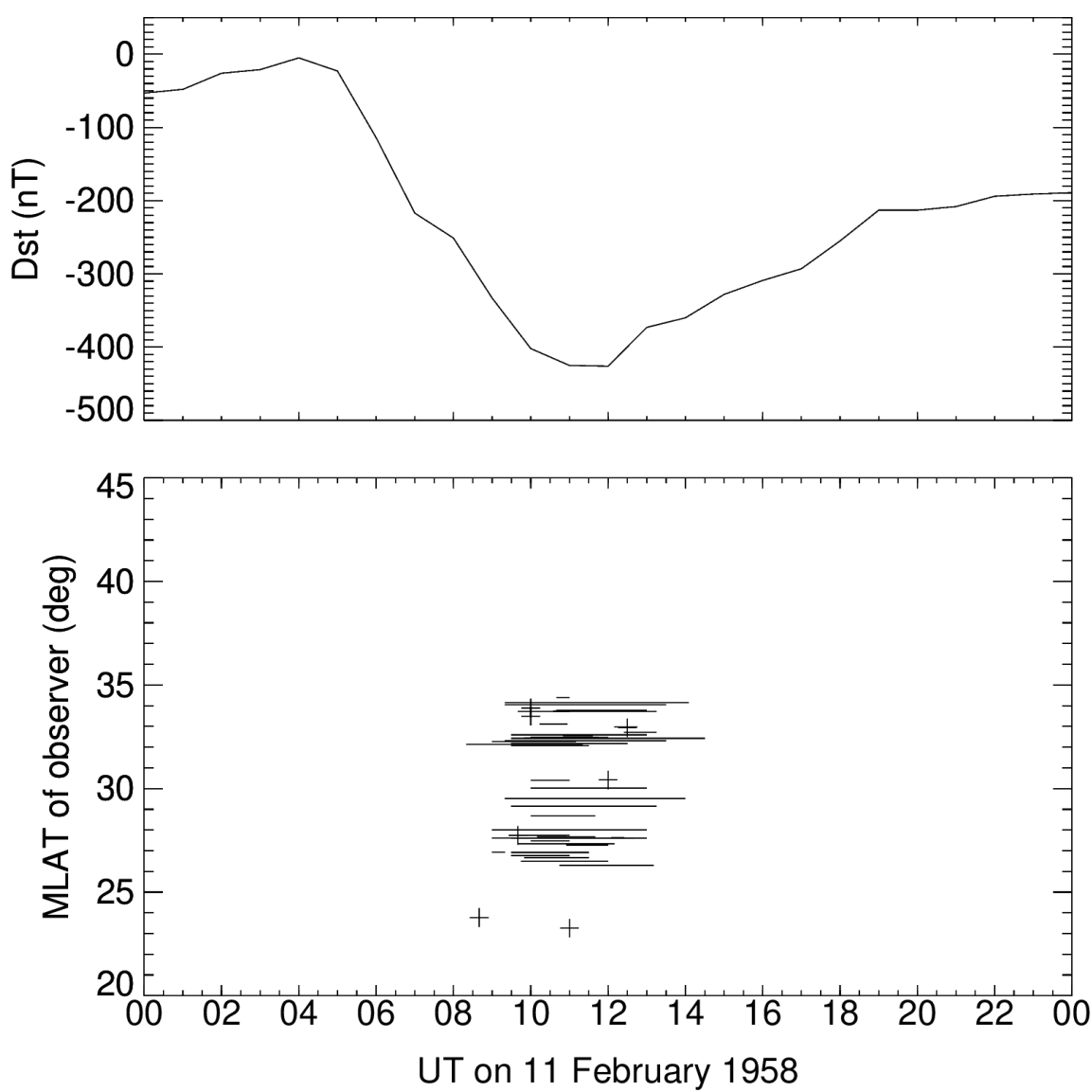

Figure 14: Temporal and spatial evolution of the auroral visibility on 11 February 1958 (lower panel), contextualised in temporal variation of the Dst index. In the lower panel, the Japanese local

time (UT + 9 h) has been corrected to the UT. When the end of the auroral visibility was neither described nor indicated, we only visualised the onset with a cross mark (+).

Figure 14 contextualises temporal evolutions of these auroral visibilities upon the Dst index (WDC for Geomagnetism at Kyoto *et al.*, 2015), representing the fourth greatest geomagnetic storm in the Dst index (minimum Dst = −426 nT), which peaked at 12 UT on 11 February. Their visibilities were reported from 17:40 – 23:05 LT in Japan (08:40 – 14:05 UT). This duration is chronologically located in the main phase to the early recovery phase of this geomagnetic storm, where the Dst index exceeded the threshold of Dst ≤ −330 nT.

**6. Isolated Auroral Reports**

In addition, these records confirm seven more nights with isolated aurorae in Japan around the IGY. Table 1 summarises their profile in terms of dates and equatorward boundaries of the auroral visibility and auroral oval, assuming the auroral elevation to be ≈ 400 km (Roach *et al.*, 1960; Ebihara *et al.*, 2017). These isolated auroral displays were reported from one or two observer(s), in contrast with the three great auroral displays attested by multiple reports (Sections 3 – 5). This was possibly because their durations were commonly short (≤ 1 hour). Without significant brightness, such short durations may have hindered sufficient attention from the ground observers.

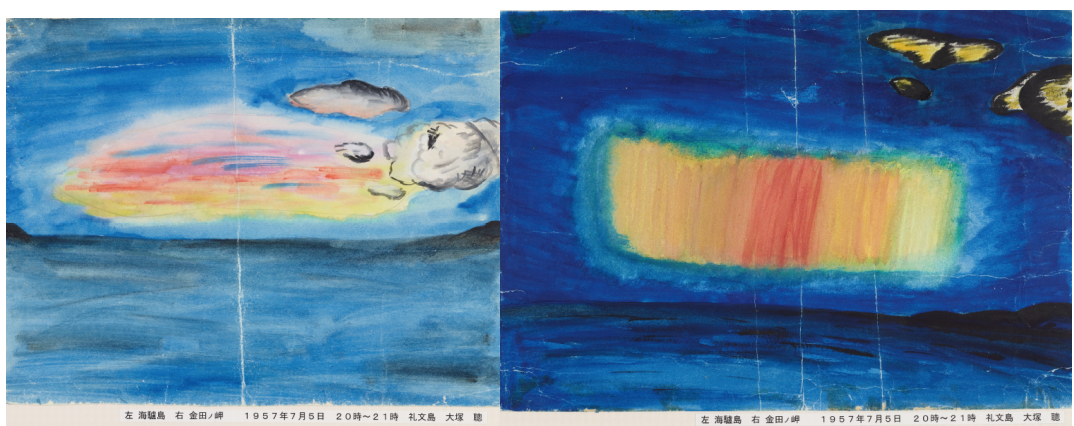

Figure 15: Examples of drawings of the auroral display on 5 July 1957 by Otsuka, reproduced courtesy of © Satoshi Otsuka and © the National Astronomical Observatory of Japan. While its duration was described as 20 – 21 LT in the image caption, its contemporary report located its onset as 20:10 LT (*e.g.*, JMA, 1957b); we opted for the duration in the contemporary report.

Still, two of the aforementioned auroral events drew moderate attentions (*e.g.*, JMA, 1957b, 1957c).

The first is an auroral display on 5/6 July 1957. Its visibility was reported from Oshonnai Village in Rebun Island (35.3° MLAT). The verbal report and coloured drawings of this event indicate its colourations as reddish, yellowish, orange, pinkish, and purplish (JMA, 1957b; Figure 15). Its contemporary report shows the visibility duration as 20:10 – 21:00 LT (11:10 – 12:00 UT), which was part of the recovery phase of a moderate geomagnetic storm on 5/6 July 1957. The second is an auroral display on 21/22 September 1957. This auroral display was reported in Suttsu (32.6° MLAT) during 22:50 – 23:08 LT (13:50 – 14:08 UT), whereas its maximal elevation was unclear owing to the contemporary cloud cover (JMA, 1957c, p. 34). Another archival report preserved at Kiso Observatory claimed auroral visibility at Takaradzuka (24.2° MLAT) during 22:40 – 23:00 LT (13:40 – 14:00 UT); however, the JMA was rather sceptical with respect to its reliability (JMA, 1957c, p. 34). This is chronologically contextualised in an early main phase of the geomagnetic storm series during 21 – 24 September 1957.

We compared these auroral records with the geomagnetic disturbances in the Dst index within ± 3 days, following Willis *et al.*'s (2007) procedures. Following Loewe and Prölss's (1960) classification, we associated two nights with severe storms (−350 nT < minimum Dst ≤ −200 nT), one night with a strong storm (−200 nT < minimum Dst ≤ −100 nT), one night with a moderate storm (−100 nT < minimum Dst ≤ −50 nT), two nights with weak storms (−50 nT < minimum Dst ≤ −30 nT), and one night with no storms (minimum Dst > −30 nT). Given their lower MLAT (≈ 30° – 36°), no significant geomagnetic storms were reported in the auroral displays on 20 June and 13 July 1957, and 17 May 1960 (within ± 3 days). Still, observational evidence shows certain auroral displays locally reported without quasi-simultaneous geomagnetic storms Japanese spectroscopic observations (Shiokawa *et al.*, 2005). Similar visual aurorae were reported in the United States without quasi-simultaneous geomagnetic storms and have been called "sporadic aurorae" (Silverman, 2003). Further analyses are required for these events and parallel cases.

Table 1: Summary of the Japanese visual auroral records around the IGY. EBAV and EVAO abbreviate the equatorward boundaries of the auroral visibility and auroral oval, respectively. The EBAO on 17 May 1960 was not estimated, as the auroral altitude was not recorded for this event.

| Year | Month | Date | EBAV (°) | EBAO (°) | min. Dst (nT) | Remarks |
|------|-------|------|----------|----------|---------------|-----------|
| 1957 | 3 | 2 | 30.1 | 40.6 | -255 | Section 3 |

| 1957 | 6 | 20 | 31.9 | 43.5 | -35  |           |
|------|---|----|------|------|------|-----------|
| 1957 | 7 | 5  | 36.2 | 42.6 | -101 | Figure 15 |
| 1957 | 7 | 6  | 33.7 | 47.4 | -92  |           |
| 1957 | 7 | 13 | 33.9 | 41.4 | -26  |           |
| 1957 | 9 | 13 | 31.9 | 38.3 | -427 | Section 4 |
| 1957 | 9 | 21 | 32.6 | 42.4 | -282 |           |
| 1958 | 2 | 11 | 23.3 | 33.3 | -426 | Section 5 |
| 1960 | 3 | 30 | 29.7 | 40.8 | -327 |           |
| 1960 | 5 | 17 | 32.9 | --   | -38  |           |

**7. Summary and Discussions**

This article reviews and envisions the Japanese visual auroral records during the IGY (1957 – 1958). Huruhata took part in this international scientific collaboration and organised an observational network throughout Japanese Islands, involving professional observatories and citizen contributors. In this regard, Huruhata's approach looks like an early prototype of space-weather citizen science projects (*e.g.*, MacDonald *et al*., 2015). These observational reports were collected to the JMA and Tokyo Astronomical Observatory. We identified these records in the contemporary JMA journals (JMA, 1957a, 1957b, 1957c, 1958a, 1958b), the original daily ledgers at the JMA Local Meteorological Offices, digital images preserved at Kiso Observatory of the University of Tokyo, and auroral drawings at the National Astronomical Observatory of Japan (formerly Tokyo Astronomical Observatory). Their details have been visualised in this study, to facilitate further analyses of the extreme space weather events and the auroral activity around the maximum of Solar Cycle 19.

These primary records have provided rich reference for the three extreme geomagnetic storms on 2/3 March 1957 (minimum Dst = −255 nT), 13/14 September 1957 (minimum Dst = −427 nT), and 11/12 February 1958 (minimum Dst = −426 nT). In Japan, the equatorward visibility boundaries of the aforementioned storms have been confirmed to extend up to Kuji (30.1° MLAT), Mori (31.9° MLAT), and Ogori (23.3° MLAT), respectively (Figures 2, 6 and 13). Based on these geomagnetic storms, we reconstructed the equatorward boundaries of these auroral ovals down to 41.4° ILAT, 38.3° ILAT, and 33.3° ILAT, respectively. The spatial extent of the great auroral display in February 1958 was significantly larger than previously considered (*e.g.*, Kataoka *et al*., 2019a, 2019b; Knipp *et al*., 2021), owing to the rich datasets from the archival collections such as multiple auroral drawings and the earliest auroral images (Figures 8 – 12). Its visibility extent was significantly larger than that in September 1957, despite their almost identical storm magnitudes. This was partially attributed to the different weather conditions and cloud covers in September 1957. These auroral records have been located on the main phase to the early recovery phase of the contemporary geomagnetic storms (Figures 3, 7, and 14).

Their spatial extents compare well with those of the extreme storms in history. During the March 1989 storm, satellite observations detected extensions of the auroral particle precipitations down to 40.1° MLAT (Rich and Denig, 1992), and this geomagnetic storm was the greatest (minimum Dst = −589 nT) since the IGY (Allen *et al*., 1989; Boteler, 2019). We have also extended the comparison with the extreme storms before the Dst index, where their intensities have been estimated with the Dst estimates (Dst*), on the basis of four reference magnetograms at the mid/low MLATs, which replace the reference stations of the standard Dst index. On their basis, the IGY storms also compare well with the great auroral displays during the extreme geomagnetic storms such as those on 25 September 1909 (EBAO ≈ 31.6° ILAT vs minimum Dst* ≈ −595 nT; Silverman, 1995; Hayakawa *et al*., 2019a), 21/22 and 25/26 January 1938 (EBAO ≈ 40.3° ILAT vs minimum Dcx ≤ −328 nT and EBAO ≈ 40.0° ILAT vs minimum Dcx ≤ −336 nT; Hayakawa *et al*., 2021b), 1 March 1941 (EBAO ≈ 38.5° ILAT vs minimum Dst* ≤ −464 nT; Hayakawa *et al*., 2021a), and 26 March 1946 (EBAO ≤ 41.8° ILAT vs minimum Dst* ≤ −512 nT; Hayakawa *et al*., 2020). Our case reports benefit further comparisons of the EBAO with the intensity of the associated geomagnetic storms (Yokoyama *et al*., 1998; Blake *et al*., 2021), as only few extreme geomagnetic storms have been subjected to analyses in these statistical studies.

Their colourations are observed to be mainly reddish and occasionally greenish, whitish, pinkish, bluish, and yellowish to orange. The coexistence of the reddish and greenish colourations indicates aurora dominated by oxygen emissions at 630.0 nm [OI] and at 557.7 nm [OI] (Tinsley *et al*., 1984), rather than a SAR arc (Kozyra *et al*., 1997). The yellowish colourations between the lower hem and the higher, red-dominated regions are explained in terms of mixture of greenish and reddish colourations (Chamberlain, 1961), or atmospheric extinction of greenish colouration (Kataoka *et al*., 2019a). The whitish colouration is typically interpreted as the greenish emissions that are not bright enough for the human eye and with possible contributions from other emissions (Ebihara *et al*., 2017; Stephenson *et al*., 2019; Bhaskar *et al*., 2020). If the pinkish colouration was horizontally narrow, and rapidly moving in the horizontal direction, it would be attributed to a ray. Narrow rays are known to exhibit a violet colour at the leading edge and a green colour behind owing to the lifetime of O($^1$S) being longer than that of excited $N_2^+$ (Omholt, 1971, pp.126). The delay of the 557.7 nm [OI] emission is confirmed by simultaneous observations of aurora and precipitating electrons (Ebihara *et al.*, 2009). The pinkish or bluish colouration extending to higher altitudes is also attributed to nitrogen emissions ($N_2^+$) in sunlit aurorae (Hunten, 2003; Shiokawa *et al.*, 2019).

In contrast, it is challenging to interpret the colouration of yellowish to orange pillars. This colouration occurs following a mixture of greenish (oxygen emissions at 557.7 nm [OI]) and reddish (oxygen emissions at 630.0 nm [OI]) colourations. The reddish emissions are, in general, dominant at high altitudes where the quenching can be disregarded (Harang, 1956; Rees *et al.*, 1967). At high altitudes, the greenish emission at 557.7 nm [OI] can increase significantly as a backtrail against the background reddish emission at 630.0 nm [OI] if electron precipitations are narrowly confined and spatially moved within the auroral display owing to the shorter lifetime of O($^1$S) ($\approx$ 0.7 s) compared to the lifetime of O($^1$D) ($\approx$ 110 s). This "lifetime hypothesis" is consistent with the contemporary descriptions of the yellowish pillars, which are reported to be pulsating every minute and are likened to a swaying curtain in a theatre (JMA, 1958a, pp. 31 – 33; Figure 12). Their width was reported to be 1.1° at Abashiri (JMA, 1958a, p. 31). If we assume the distance from the observational site to be 400 km, their width is computed as 7.7 km. If the electron precipitation occurred in a considerably narrow region (much shorter than 7.7 km), and the brightness increases and decreases with a lifetime of 0.7 s (Klekociuk and Burns, 1995), we could estimate the east-westward speed as 6 km/s. The estimated speed is lower than that of the rays found in active aurorae (Omholt, 1962). This is also significantly faster than their westward propagation of 0.4 km/s at the altitude of 400 km, derived from spectroscopic observations at Memanbetsu (Kataoka *et al.*, 2019a). The exposure time of 7 s of the photograph is probably insufficient to resolve the individual rays moving with $\approx$ 6 km/s. The pillars moving with $\approx$ 0.4 km would result in an aftertrail of 557.7 nm [OI] with thickness of 0.56 km. The thickness of 0.56 km could be too small to be sufficiently resolved by the photograph (Kataoka *et al.*, 2019a). It is likely that the slowly moving pillars captured by the photograph manifest bulk motion of the rapidly moving rays. The spectroscopic observations show that the intensity of the 630.0 nm [OI] is an order of magnitude larger than that of 557.7 nm [OI] (Kataoka *et al.*, 2019a). The exposure time (35 – 60 min) of the spectroscopic observations is also too long to resolve the aftertrail in which the greenish emission 557.7 nm [OI] could dominate the reddish one 630.0 nm [OI]. Kataoka *et al.* (2019a) attributed the white pillars to be the dominant green line, with

Hayakawa et al. (2021) Japanese auroral records in the IGY (1957 – 1958)
*Geoscience Data Journal*, DOI: 10.1002/GDJ3.140a small contribution from blue line. The "lifetime hypothesis" reasonably accounts for the rays dominated by the greenish emission 557.7 nm [OI] and the descriptions of yellowish pillars extending higher altitudes.

Overall, these visual auroral reports provide unique details on the low-latitude aurorae for the three severe geomagnetic storms, two of which exceed the threshold of minimum Dst = −400 nT, and on the visual auroral activities in the low-latitude region (20° − 35° MLAT) around the greatest solar cycle since 1610. These analogue reports bridge our knowledge on the extreme space weather events from the space age to the pre-IGY age. It would be beneficial to further investigate the visual auroral reports around the IGY for the comprehensive reconstruction of low-latitude aurorae on a global scale.

**Acknowledgment**

We first thank the individual observers for their observations, Masaaki Huruhata for his effort to collect these reports, and Kyoko Tanaka for her effort for their preservation. We thank Mizuho Tamefusa for letting us access the 'Sketches for aurorae that occurred during the International Geophysical Year' at the National Astronomical Observatory of Japan (NAOJ) Library, Tsutomu Aoki, Yuki Mori, and Naoto Kobayashi for helping us to access digital images preserved at Kiso Observatory of the University of Tokyo, and the Local Meteorological Offices at Niigata and Sapporo for letting us access their contemporary original daily ledgers. We consulted the contemporary weather map through the Database of Weather Charts for Hundred Years[2]. We thank Kohji Ohnishi for connecting us with staffs of Kiso Observatory of the University of Tokyo, and Hiromi Kawai, Ryuma Daigo, Takuya Takahashi, Machi Akida, Kotaro Araki, Takanori Ito, and Ikuko Sato for their help on accessing several contemporary publications. HH has benefited from discussions within the ISSI International Team #510 (SEESUP Solar Extreme Events: Setting Up a Paradigm) and ISWAT-COSPAR S1-01 and S1-02 teams.

**Funding Information**

This work was financially supported in part by JSPS Grant-in-Aids JP20K22367, JP20K20918, JP20H05643, and JP21K13957, JSPS Overseas Challenge Program for Young Researchers, Otaru Studies of Otaru University of Commerce, Comprehensive Research on the Slavic Eurasian Region---

[2] http://agora.ex.nii.ac.jp/digital-typhoon/weather-chart/index.html

**Conflicts of Interests**

The authors have declared no conflicts if interests.

Hayakawa et al. (2021) Japanese auroral records in the IGY (1957 – 1958)
*Geoscience Data Journal*, DOI: 10.1002/GDJ3.140